\documentclass[conference]{IEEEtran}
\usepackage{graphicx}
\usepackage{multirow}
\usepackage{amssymb}
\usepackage{amsmath}
\usepackage{subfigure}
\usepackage{url}
\usepackage{caption}
\usepackage{color}
\usepackage{amsmath}
\usepackage{sidecap}
\usepackage{cite}
\usepackage{array}
\ifCLASSINFOpdf

\else
\fi

\begin{document}
%
\title{An Automated System for Discovering Neighborhood Patterns in Ego Networks}

\author{\IEEEauthorblockN{Syed Agha Muhammad}
\IEEEauthorblockA{Embedded Sensing Systems\\
Technische Universit\"at Darmstadt\\
Email: aghahashmi@gmail.com}
\and
\IEEEauthorblockN{Kristof Van Laerhoven}
\IEEEauthorblockA{Embedded Systems\\
University of Freiburg\\
Email: kristof@ese.uni-freiburg.de}}

\maketitle

\begin{abstract}
Generally, social network analysis has often focused on the topology of the network without considering the characteristics of individuals involved in them. Less attention is given to study the behavior of individuals, considering they are the basic entity of a graph. Given a mobile social network graph, what are good features to extract key information from the nodes?
How many distinct neighborhood patterns exist for ego nodes? What clues does such information provide to study nodes over a
long period of time? 

In this report, we develop an automated system in order to discover the occurrences of prototypical ego-centric patterns from data. We aim to provide a data-driven instrument to be used in behavioral sciences for graph interpretations. We analyze social networks derived from real-world data collected with smart-phones. We select 13 well-known network measures, especially those concerned with ego graphs. We form eight feature subsets and then assess their performance using unsupervised clustering techniques to discover distinguishing ego-centric patterns. From clustering analysis, we discover that eight distinct neighborhood patterns have emerged. This categorization allows concise analysis of users' data as they change over time. The results provide a fine-grained analysis for the contribution of different feature sets to detect unique clustering patterns.  Last, as a case study, two datasets are studied over long periods to demonstrate the utility of this method. The study shows the effectiveness of the proposed approach in discovering important trends from data. This is a technical report for our paper \cite{ess_2015}.

 
\end{abstract}

\section{Introduction}

Social network analysis has acquired enormous popularity and has become a key determinant in modern sociology.  It is now widely used in behavioral science, psychology, biology, history etc. The main component of a social network is an actor, also known as an ego node, that shares social ties and relationships with other actors. While earlier analysis on ego networks focused on studying simple graph properties such as degree distribution, diameter, in and out degree of nodes and simple graph patterns such as clique, more recent methods aim at detecting complex neighborhood circles \cite{22} in the vicinity of an ego network. The direct proximity network of an individual provides an enormous amount of information on the possible relationships, social circles and intimacy levels with its neighbors, and in addition, it provides characteristic information about an individual himself.  

Traditionally, social network data was gathered using survey based methods, where an observer was planted to collect the data. Nowadays, our mobile phones are equipped with different types of built-in sensors, such as Bluetooth, GPS, WiFi, call-log, application usage, etc. The growing utilization of mobile phones has led to an interest in gathering the obstructive social interaction data for a variety of practical applications. The low threshold and cost effective means of capturing such a data provides an ample scientific opportunity to study the structure and dynamics of large social networks at different levels; starting from the small-scale individual patterns to the large scale collective group behaviors, with an unprecedented degree of reach and accuracy.

\begin{figure}[ht]
\centering
 \begin{subfigure}[Star graph with a single triangle]{\includegraphics[trim = 0mm 0mm 0mm 30mm,scale=.20]{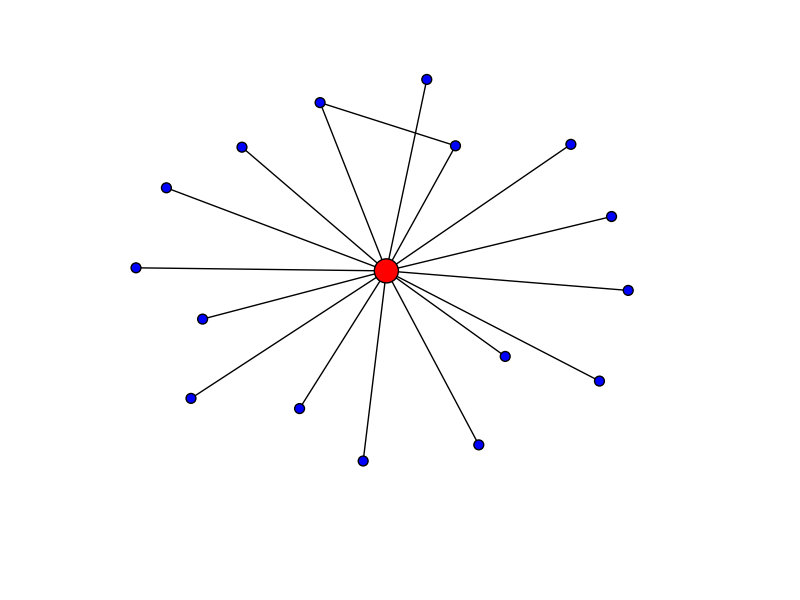}
   \label{fig:subfig4}
 }%
\end{subfigure}\hfill
 \begin{subfigure}[Star Graph with multiple triangles]{\includegraphics[trim = 0mm 0mm 0mm 30mm,scale=.20]{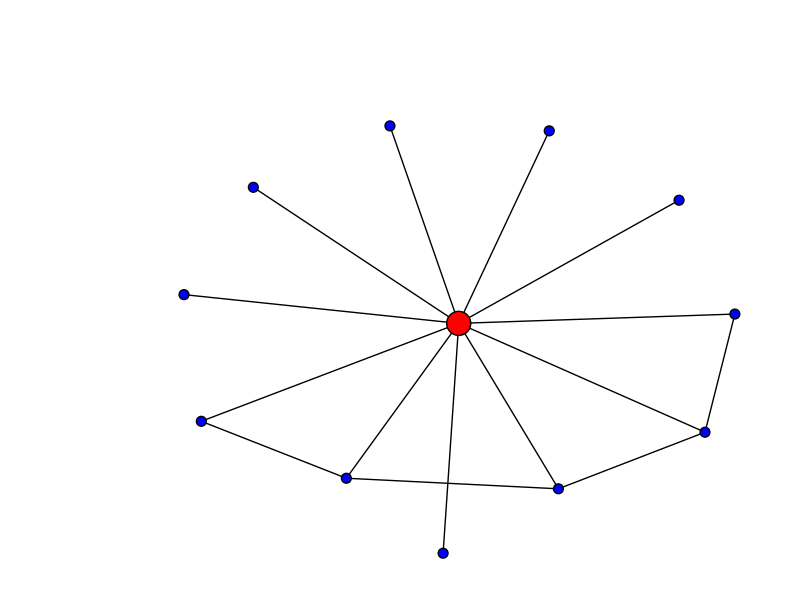}
   \label{fig:subfig6}
 }%
\end{subfigure}%
\label{egop}
\caption{Ego graphs, depicting the ego node in the middle.}
\end{figure}
  
{\bf Overview of the Approach.} In this report, we explore ego-centric patterns to discover the occurrences of most prototypical neighborhood patterns, understanding the different types of possible human interactions and how they evolve over a time. We design an automated system to detect the distinguishing ego-centric patterns from data. The work aims to provide a data-driven instrument to be used in behavioral sciences for graph interpretations. In sociological studies, the neighborhood patterns can be interpreted based on different aspects, such as cultural norms, environmental conditions, etc. To do so, we consider a set of four spatial-temporal datasets; 13 network level features to extract key attributes from ego networks, and three different unsupervised clustering algorithms to detect clusters from data. One should note that we are primarily interested in detecting different ego-centric patterns from data, and then examining how these patterns morph with time. 
\begin{figure*}[!htb]
\centering

\begin{center}
\includegraphics[trim = 0mm 30mm 00mm 40mm,scale=.35,clip=true]{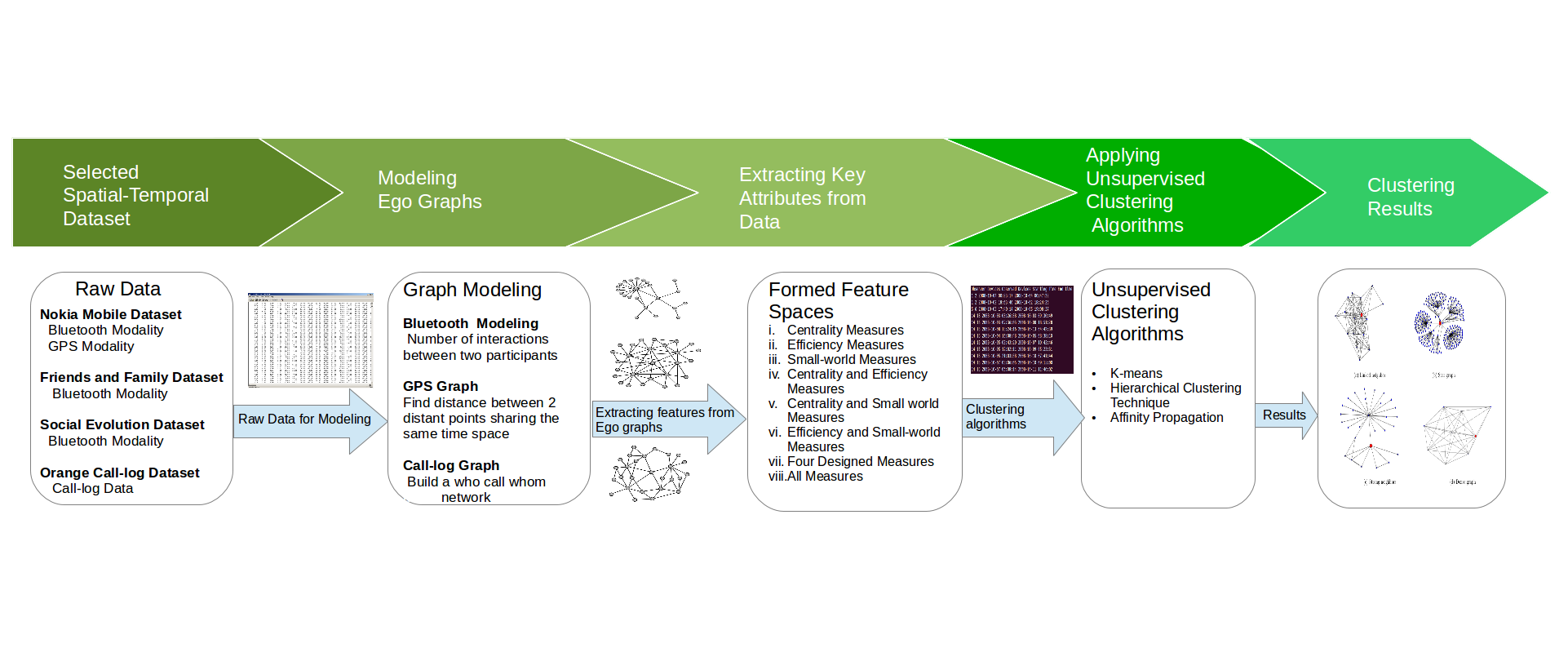}
\end{center}
\caption{Overview of our method. Initially, we select different modalities Bluetooth, GPS, and call-log data from four datasets. In the next phase, weighted graphs are  modeled for each modality. After that different feature spaces are created and key attributes are derived from ego graphs. To cluster ego graphs, different unsupervised clustering algorithms are selected. In the last phase, clustering results are evaluated based on their shapes. }
\label{fig1}
\end{figure*}
Figure 1 illustrates a basic understanding of our approach. From the outset, one may suggest two unique patterns; however, the trends are similar in both graphs which is star patterns and very few connected components. We model social networks from smart-phones collected data, i.e. Bluetooth, GPS,  call-log data. We use four centrality measures (degree, betweenness, closeness, and eigenvector centrality), three efficiency measures (global, local and nodal efficiency), three transitivity measures (global and local), and four designed measures (ego density, ego neighbors, dominant edges in an ego network, and ego weight). We perform exhaustive search and find eight different feature subspaces from the above mentioned network measures and then apply clustering algorithms to detect ego-centric patterns from data. For clustering purposes, we use k-means, hierarchical and affinity propagation techniques.  
We create eight different feature subsets and then apply clustering algorithms to detect neighborhood patterns. We examine the properties of the delivered clusters by different feature subsets and algorithms to demonstrate how their combinations produce different neighborhood patterns and which combinations produce the optimal number of clusters. We perform case studies on two datasets for several months as a proof of concept. The selected datasets have the ground-truth data of the participants. Our results have shown that the availability of such a tool can be facilitating to study ego networks over a long period of time. Figure~\ref{fig1} illustrates an overview of our approach. To the best of our knowledge, this is the first work that explores various graph measures on spatial-temporal data to automatically infer distinct neighborhood patterns from ego graphs. A large deal of work has been devoted on inferring five personality traits from social networks\cite{1,2,3,20,28}, detecting Dunnbar inner circles in ego graphs \cite{23,24}, comparing ego-centric data from social networks and spatial-temporal data \cite{21}, and discovering social circles from social networking data \cite{22}. Within the graph mining community, the approaches developed till now have focused on detecting frequent subgraphs patterns\cite{13,19,5} and top-k graph patterns\cite{9,10}. There exists no graph mining technique to detect ego graph patterns from the data.

The remainder of the report is organized as follows: Section II discusses background and related work on ego networks. We describe our methodology in Section III. We discuss the datasets in Section IV. In Section V, we present our experimental results. In Section VI, we present our case study. We conclude the report in Section VII.
\section{Background and Related Work}
In this section, we discuss the network features used in this work and the summarize the related work. 

\subsection{Background}
An undirected graph is denoted by $G(V,E)$ where $V$ is a set of nodes and $E$ is a set of edges. Given a graph $G$ and a node $v\in V$, Ego Network ($v, G$) is a sub-graph $\hat G$($\hat V$,$\hat E$), where $\hat V$ represents the direct neighbors of $v$ and $\hat E$ represents all its neighbors in $E$. In this work, we explore the first and second order neighborhood of an ego network. In the remainder of this section, we discuss the network features. Table I summarizes the used network features.  
\begin{table}[!ht]
\small
\begin{center}
    \begin{tabular}{ | p{3cm}| p{4cm} |}
    \hline
    { \bf Feature Category} & { \bf Selected Features}  \\ \hline
    Centrality Measures & Degree, Betweenness, Closeness, Eigenvector   \\ \hline
    Efficiency Measures & Global, Local, Nodal \\\hline
    Transitivity Measures &  Global, Local \\\hline
    Actor Based Measures &  Ego Density, Ego Neighbors, Dominant Edges, Ego Weight \\\hline
    \end{tabular}
\end{center}
\caption{\small{Extracted network features for ego graphs.}}
\label{turns}
\end{table}

{\bf Centrality Measures}
There are different centrality measures available in literature, but most famous amongst them are degree, betweenness and closeness centrality \cite{7}. 

Degree centrality is the number of edges incident upon a node, with which it is in direct contact. Any node, whose position allows it to be in a direct contact with many other nodes is perceived to be a major channel of communication. Closeness centrality represents the degree to which a node is close to other nodes in the network. Normally, a higher closeness suggests the capability of a node to send information quickly across its neighbors.

The two measures mentioned above are directly based on how close the ego node is to the other nodes, while betweenness centrality is based on the geodesic distance between the specific node and the remaining nodes. It represents the frequency with which a node is placed between two nodes on the shortest path connecting them. A node occupying such a central position controls the communication in the network. Another centrality measure we extract, eigenvector centrality, adds a centrality score to each node depending on degree and weight or quality of its edges. 

{\bf Efficiency Measures}
It measures \cite{8} how efficiently information is exchanged over the network, and to characterize the closeness of the ego to the small-world model.

Small world networks are special kinds of networks where most of the nodes are not direct neighbors to each other, but where most nodes can be reached by a small number of steps. Small world networks are highly clustered, and have small characteristic paths like random graphs. The global efficiency of the network $E(G)$ containing nodes $N$ is defined as: 
\begin{equation}
	\small
	E(G) = \frac{1}{N(N-1)} \sum_{i\not=j\epsilon G} \frac{1}{d_{ij}}
\end{equation}
$d_{ij}$ denotes the shortest path length between $i$ and $j$, and $1/N(N-1)$ is the normalizing factor, with a value between 0 and 1, where 1 represents high and 0 represents low efficiency. 
For each node $i$ in the graph, the local efficiency is defined as 
\begin{equation}
	\small
	E_{loc} = \frac{1}{N}\sum_{i\not=j\epsilon G} \frac{E(G_i)}{({G_{i}^{ideal}})}
\end{equation}
where for each node $i$, $E({G_{i}^{ideal}})$ is the efficiency of the ideal case, when $G_{i}$ has all possible $k_{i}$($k_{i}$-1)/2 edges, where $k_{i}$ represents the edges incident with $i$. The local efficiency describes how fault tolerant the system is, which means in case when node $i$ is removed from the graph, how efficient the communication between the first neighbors of $i$ remains. Higher values of global and local efficiency suggest a model which is nearer to a small-world model. 

Nodal efficiency of a node $i$ is defined as the inverse of the harmonic mean of the path length. We extracted nodal and local efficiency for each node in the networks. For a node $i\in G$, it can be calculated as:
\begin{equation}
	\small
	E^{nodal}_i = \frac{1}{(N-1)} \sum_{i\epsilon G} \frac{1}{L_{ij}}
\end{equation}

{\bf Transitivity Measures}
Transitivity measures the probability that a neighborhood of an individual node is connected. It uses the concept of triple, which is a set of three nodes, that can be closely connected to each other (close loop), or two out of three nodes are connected (open triple). The global transitivity of a given graph $G$ is the ratio between the number of closed triples in $G$ and the total number of triples. It gives clustering value at the level of the entire graph. Normally, for global transitivity the number of triples are counted in the ego graph. For each ego, presence of triples gives an indication of the clustering in a network. The local transitivity of a node measures how concentrated its neighbors are to forming a clique and the graph to a small-world network. 

{\bf Actor Based Measures}
Apart from the above mentioned features, we also include some additional features.  

The density of an ego node $\hat G$ is calculated as:
\begin{equation}
	\small
  Ego Density (\hat G) = \frac{2 \times|E|}{|N| \times|N-1|}
\end{equation}
The neighborhood of an ego node $v$ is calculated as: 
\begin{equation}
	\small
	Ego Neighboors (\hat G, v) = \sum_{j\epsilon \hat G(v)}{k_{vj}}
\end{equation}
We design a feature to detect the dominant edges from ego graphs. The dominant edges in $\hat G$ are calculated as: 
\begin{equation}
	\small  
  D\_Edges(\hat G)= \log{(\max({EW- mean(EW)}))}
\end{equation}
where $EW$ represents the weight of each edge in $\hat G$.

Similarly, we calculate the weight of an ego graph by adding the weight of each edge in it. 

\subsection{Related Work}

The problem of studying ego networks is a crucial problem and there is a variety of research conducted to formalize different aspects of this issue. A considerable amount of attention is devoted on studying five-personality traits from survey, spatial-temporal, and web mining data. Staiano et al. \cite{3} used network level features on a spatial-temporal dataset collected in an undergraduate student campus to investigate personality traits. Their research shows that Bluetooth data identified different personality traits much better than call-log, survey, Bluetooth and call-log data together. Chittaranjan et al. \cite{2} developed an automated system for classifying personality traits based on actor level features, such as the use of camera, application, youtube videos, incoming call duration, Bluetooth information, etc. Wahrli et al. \cite{1} predicted personality traits from social networking website StudiVz using network and actor based features. Pan et al. \cite{28} studied spatial-temporal and survey based data of ego networks to identify the existence of individual-level correlation between financial status and interaction patterns, and their connection to personality traits.

Apart from the five-traits model, other models are also developed for studying ego networks. Stocio et al. \cite{4} presented a model to characterize undirected graphs by enumerating small induced sub-structures. Socievole et al. \cite{21} analyzed the Bluetooth and social networking data from a particular group of people using socio-centric measures (betweenness, closeness, eigenvector centrality and Bonacich power) and ego-centric measures (degree centrality) to highlight the structural similarities and differences between the two network types. McAuley et al. \cite{22} developed a probabilistic model to infer social circles (friends, family, college friends) from social network data. Arnaboldi et al. \cite{23,24} analyzed twitter network of 500 people to identify the social circles within the ego networks. Henderson et al. \cite{25} proposed a feature extraction model that recursively combines local (in and out degree, total degree) and neighborhood (number of within-egonet edges, and the number of edges entering and leaving the ego net) features to produce behavioral information. 
Ego networks have also been studied in the health-care domain. Madan et al. \cite{26} analyzed Bluetooth proximity scans, WiFi scans, calling, SMS networks, self-reported diet, exercise and weight-related information collected periodically over a nine-months period. Malley et al. \cite{27} discussed relationships between different network features and how these properties can be studied together in a health-care domain. 
  
\section{Methodology}

We now discuss our methodology for extracting ego clusters.

{\bf Feature selection}. We turn now to investigating the predictive power of different features discussed in Section II by creating different feature spaces, applying clustering algorithms and evaluating the delivered clustering results. We perform exhaustive search and select eight features that produced comparatively better results. For the purpose of analysis eight feature subsets were created and compared: i) centrality measures; ii) efficiency measures; iii) transitivity measures; iv) centrality and efficiency measures  -i.e. the union of i) and ii); v) centrality and transitivity measures -i.e. the union of i) and iii); vi) efficiency and transitivity measures - i.e. the union of ii) and iii); vii) four actor based features; viii) combination of all 13 measures. We also use FSFS (feature selection using feature similarity) \cite{Mitra:2002:UFS:507475.507477} to automatically select feature subsets based on the characteristics of the data.

{\bf Feature evaluation.} Afterwards, we use entropy \cite{Mitra:2002:UFS:507475.507477} and representation entropy measures to evaluate the effectiveness of the feature subsets. Basically, we minimize the likelihood for the inclusion of any spurious feature. The entropy can be defined as:      
\begin{equation}
	\small
  E =   \sum^N_{i=1} \sum^N_{j=1}(s_{ij}.log(s_{ij})+(1-s_{ij}).log(1-s_{ij}))
  \end{equation}
where 
\begin{equation}
	\small
  s_{ij} =  e^{-\alpha. dist_{ij}}
  \end{equation}
and
\begin{equation}
	\small
  \alpha =   \frac{-log(0.5)} {\overline{dist}}
  \end{equation}
Here $dist_{ij}$ is the Euclidean distance between data items $i$ and $j$ for a given feature subspace and $\overline{dist}$ is the mean dissimilarity between items in the data set for a given feature subspace.     

Let the eigenvalues of $m$ x $m$ covariance matrix of a feature set of size $m$ be $\lambda_j$, j=1,..., m, Let
\begin{equation}
	\hat{\lambda_j} = \frac{\lambda_j} {\sum^d_{j=1}\lambda_j}
\end{equation}
where 0$\leq \hat{\lambda_j}\leq$1 and ${\sum^d_{j=1}\lambda_j}$ =1. Hence, a representational entropy can be defined as:
\begin{equation}
	H_R = - \sum^d_{j=1} \hat{\lambda_j} log \hat{\lambda_j}
\end{equation}
where we expect that the final reduced feature sets have low redundancy, i.e., a high representation entropy $H^s_R$.

{\bf Unsupervised clustering algorithms.} To perform clustering, we select three well-known standard unsupervised clustering algorithms: i) hierarchical clustering, ii) $k$-means clustering, and iii) affinity propagation (AP). We select different clustering algorithms to find out the best clustering algorithm that detects the optimal distinctive clusters. Hierarchical clustering techniques are deterministic and promise good success in detecting a reasonable number of clusters. The agglomerative clustering algorithm proceeds by iteratively merging small clusters into larger ones. Similarly, $k$-means is a method commonly used to partition dataset into \textit{k} groups. It assumes \textit{k} as a fixed priori. Initially, it chooses randomly \textit{k} distinct points and then assigns each observation to its nearest centroid. The process is repeated until all the observations are assigned. The first step ends when no observation remains in pending. At this stage, the centroids are updated to be the mean of their constituent instances. The procedure is repeated until no centroids can be moved further. For the previously mentioned clustering approaches, the number of clusters needs to be predetermined; however, affinity propagation does not require the number of clusters to be determined before running it.
 
{\bf Data normalization and dimensionality reduction.} In our case, the input feature spaces are high-dimensional. The performance of most clustering algorithms tends to scale poorly on high dimensional data. For this reason, we select principle component analysis (PCA) for the dimensionality reduction. We normalized the features using equation~\ref{asf} prior to applying it for dimensionality reduction. 
\begin{equation}
	\small
   V^{*} = \frac{V-min(V)}{max(V)-min(V)}
\label{asf}
\end{equation}
In this equation, $V$ denotes the variable that is normalized, $min$ and $max$ indicate the two extremes of the variable. 

{\bf Detecting optimal number of clusters.} To identify the optimal number of clusters, we select L-method \cite{final} and gap statistic \cite{Tibshirani00estimatingthe}. This step aims at identifying the clusters that are well separated, while penalizing an increasing number of clusters.  The L-method was chosen, due to its efficiency and good performance, for hierarchical clustering algorithm. The method automatically identifies the "knee" in a 'number of clusters vs evaluation metric' chart. The knee of the chart can be interpreted as a point of transition from a high to low gain in cluster separation for an increasing number of clusters. Thus the knee indicates the ideal cluster separation between the number of clusters and the evaluation metric. Similarly, gap statistic standardizes the graph of $\log(W_k)$ ($W_k$ represents within cluster dispersion) by comparing it with a null reference distribution of the data, i.e. a distribution with no obvious clustering. The optimal number of clusters is then the value of $K$ for which $\log W_k$ falls the farthest below this reference curve. This information is contained in the following formula for the gap statistic:
\begin{equation}
	\small
   Gap_n(k) = E^*_n \{log W_k\}- log W_k
\end{equation}
We generate the reference datasets by sampling uniformly from the original dataset's. To obtain the estimate $E_n^*\{\log W_k\}$ we compute the average of B copies $\log W^*_k$, each of which is generated with a Monte Carlo sample from the reference distribution. Those $\log W^*_k$ from the B Monte Carlo replicates exhibit a standard deviation $\mathrm{sd}(k)$ which, accounting for the simulation error, is turned into the quantity:
\begin{equation}
	\small
 s_k = \sqrt{2/B}\ sd(k)
\end{equation}
Finally, the optimal number of clusters $K$ is the smallest $k$ such that $\mathrm{Gap}(k) \geq \mathrm{Gap}(k+1) - s_{k+1}$. Apart from them, we rely also on visual analysis of the low dimensionality data to decide the best possible clusters.

\section{Datasets}
We have selected five publicly available spatial-temporal datasets. We use Bluetooth, GPS and call-log data from spatial-temporal datasets. We built a call network from call-log data, where participants act as nodes and the number of calls between two nodes as edge weights. Similarly, Bluetooth and GPS networks were built with participants as nodes and the count of social interactions derived from Bluetooth and GPS as edge weights. Spatial-temporal datasets often contain noisy edges or so called 'familiar strangers' in the data. There are several techniques to prune out irrelevant edges. Thresholding is one of the popular techniques, but it is a one-size-fits-all solution, i.e. an edge may be relevant even with a low weight, because it may be the strongest possible link between two 'weak' nodes. We use \cite{serrano2009extracting} to select the relevant edges from the data. Table~\ref{t} summarizes the basic statistics of the datasets. All our graphs are undirected and weighted. Below we briefly discuss the datasets used in this work.

\begin{table}[!ht]
\small
\begin{center}
    \begin{tabular}{ | p{4cm}| c |c|c|c}
    \hline
    { \bf Network} & { \bf $|N|$}&{ \bf $ |E|$} &{ \bf   $\overline k$}   \\ \hline
    The Nokia dataset &36 & 147& 8.17   \\ \hline
    The Friends and Family dataset&40 &501&24.75 \\\hline   
    The Social Evolution dataset&74 &2,526& 68.27\\\hline
    The Orange dataset & 4,357 & 25,9110& 59.06 \\\hline
    \end{tabular}
\end{center}
\caption{\small{Basic statistics of the networks studied.}}
\label{t}
\end{table}
\subsection{The Nokia Mobile Dataset}
The Nokia dataset \cite{17} representing the spatial-temporal data of the 36 participants gathered between October 2009 and September 2011 from the French region in Switzerland. The dataset contains a wide range of behavioral data, such as Bluetooth, WiFi, GPS, Accelerometer, etc. We use Bluetooth and GPS data of the participants for graph modeling.  
\subsection{The Social Evolution Dataset}
The MIT's Social Evolution dataset \cite{16} contains the data gathered between October 2008 and May 2009 from 74 participants living in a  dormitory. The dataset contains scanned Bluetooth devices, WiFi access-points, logged call records, and SMS messages. In addition, survey experiments were designed to study the adoption of political opinions, diet, exercise, obesity, epidemiological contagion, depression and stress, dorm political issues and interpersonal relationships. 
\subsection{The Friends and Family Dataset}
The Friends and Family dataset \cite{18} contains the data collected between October 2010 and March 2011 from 40 individuals living in a married graduate student residence. The collected data has the Bluetooth, SMS and voice call data of the participants. Apart from that, monthly surveys include personality traits information, participants intimacy with each other, married couples and their children. We selected Bluetooth data of the participants to model their ego networks. The Social Evolution and this dataset is extensively used in this work to study the clustering patterns over several months.
\subsection{The Orange Dataset}
The Orange dataset \cite{6} has the ego networks of 5,000 mobile users collected in Ivory Coast by French Telecom between December 2011 and April 2012. The dataset contains the call data (source and destination) for mobile phone users, and contains first and second order neighborhood of the ego.

\section{Evaluating Clustering Results}

In this section, we discuss the clustering results from algorithms applied on the feature subsets discussed previously. We detect the occurrence of most prototypical clustering patterns from the datasets. For the given datasets, the applied gap statistic and L-method identified different number of possible clusters for different feature subsets. However, the optimal number of clusters identified by any combination of features are no more than eight. We found in total eight distinct ego graph patterns as shown in Figure 3. The average Silhouette width varies for the clusters between 0.88 and 0.38. The detected prototypical clusters have the following properties and characteristics:

\begin{figure*}

\begin{subfigure}[b][\small{Linked neighbors (C1)}]{\includegraphics[trim = 10mm 0mm 0.2mm 20mm,scale=.20] {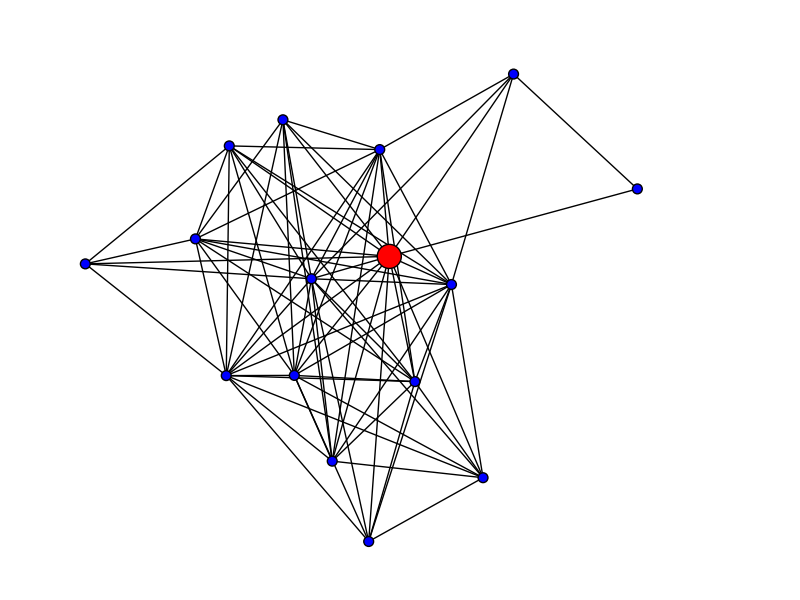}
   \label{fig:subfig1}
 }%
\end{subfigure}
 \begin{subfigure}[b][\small{Star (C2)}]{\includegraphics[trim = 20mm 0mm 0mm 20mm,scale=.20]{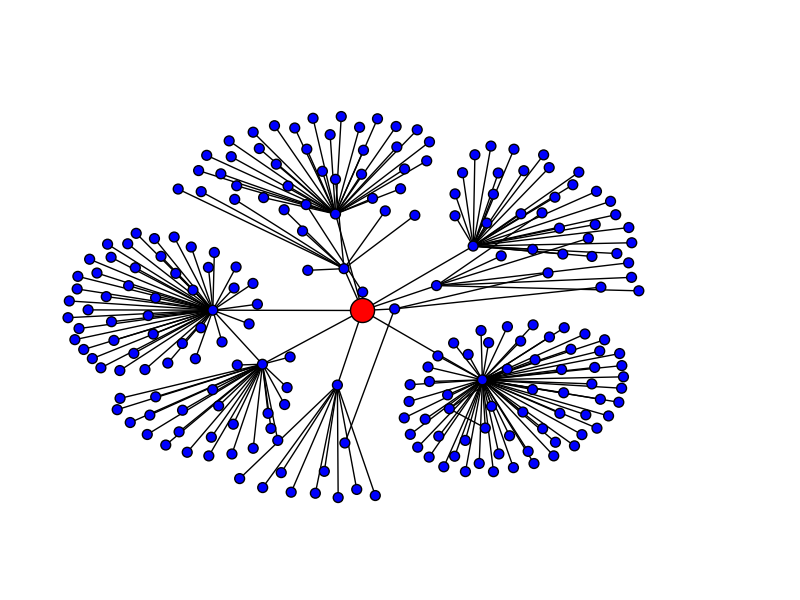}
   \label{fig:subfig2}
 }%
\end{subfigure}
 \begin{subfigure}[b][\small{Strong ego neighbor (C3)}]{\includegraphics[trim = 0mm 0mm 0mm 20mm,scale=.20]{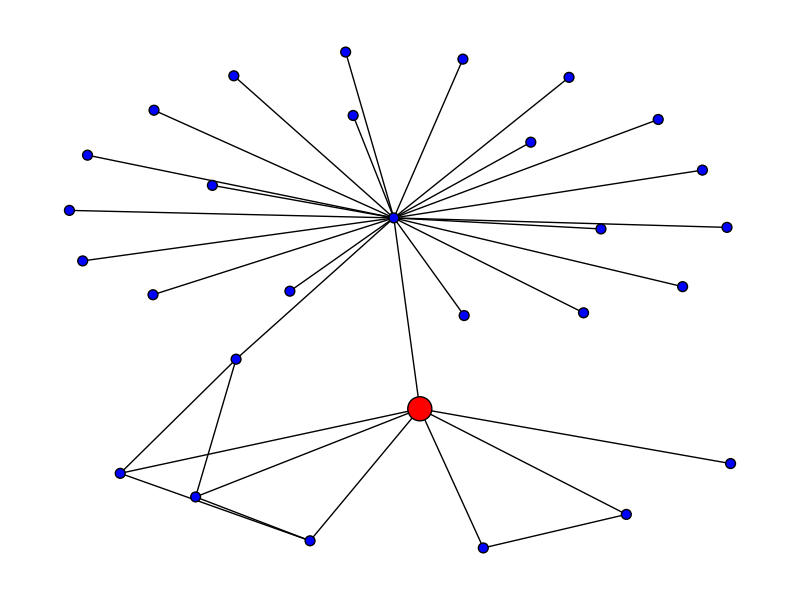}
   \label{fig:subfig3}
 }%
\end{subfigure}
 \begin{subfigure}[b][\small{Dense (C4)}]{\includegraphics[trim = 0mm 0mm 0mm 20mm,scale=.15]{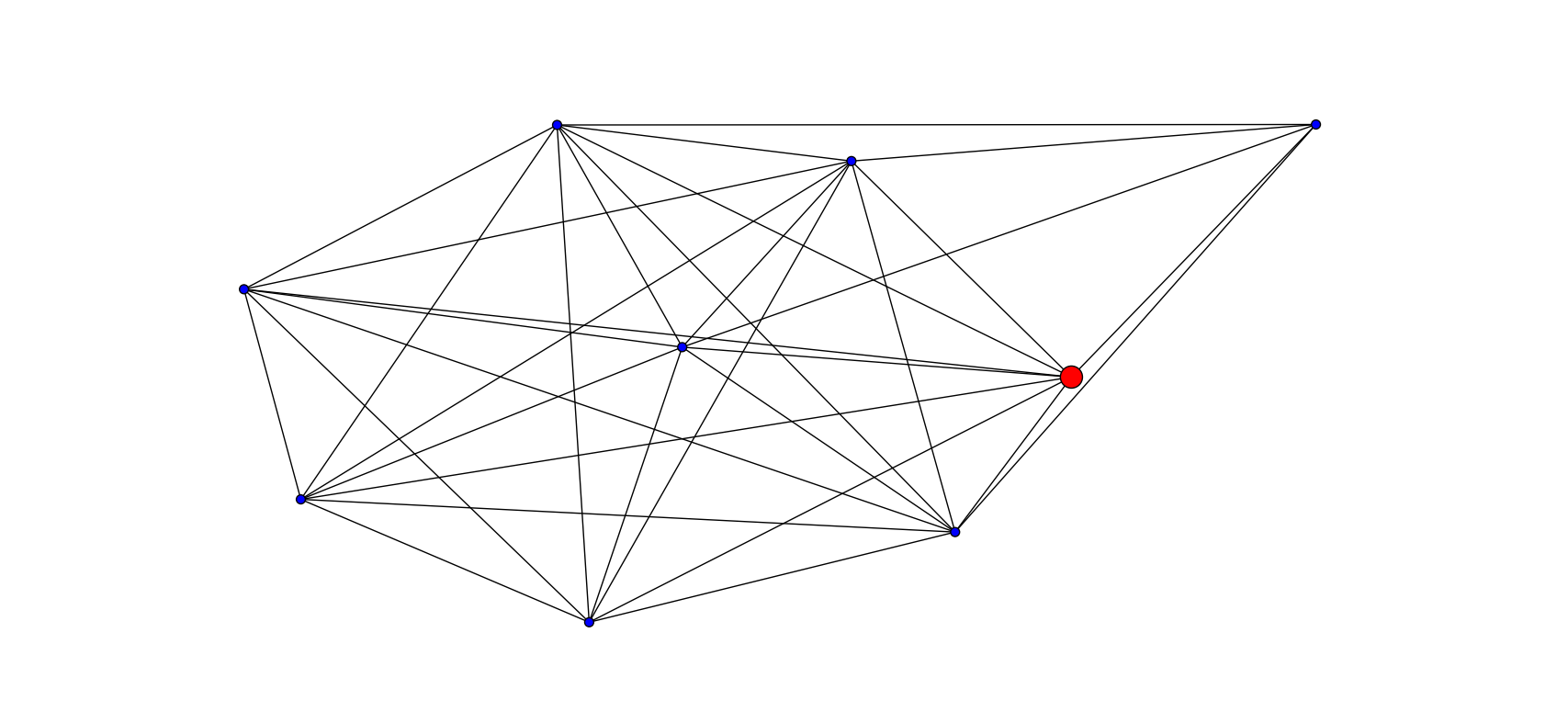}
   \label{fig:subfig4}
 }%
\end{subfigure}
\begin{subfigure}[b][\small{Powerful ego node (C5)}]{\includegraphics[trim = 10mm 0mm 0mm 10mm,scale=.20] {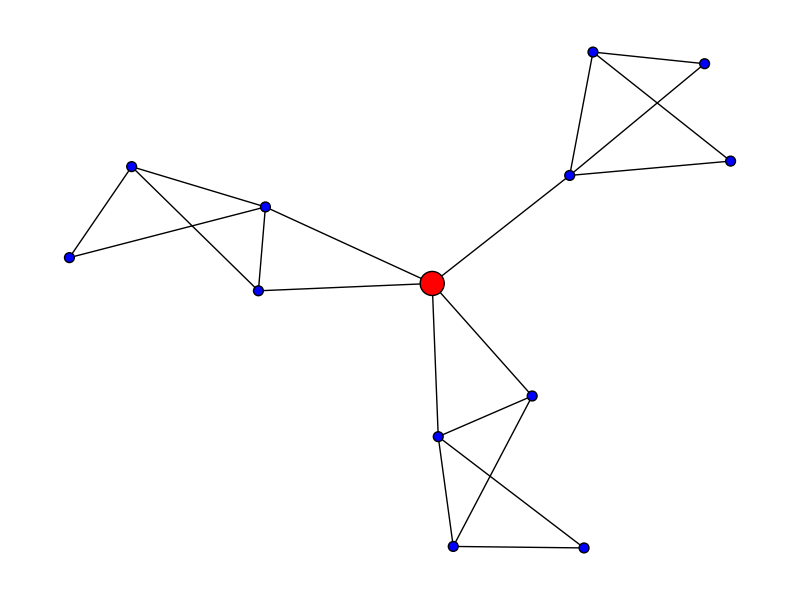}
   \label{fig:subfig5}
 }%
\end{subfigure}
 \begin{subfigure}[b][\small{Less cohesive star (C6)}]{\includegraphics[trim = 20mm 0mm 0mm 10mm,scale=.20]{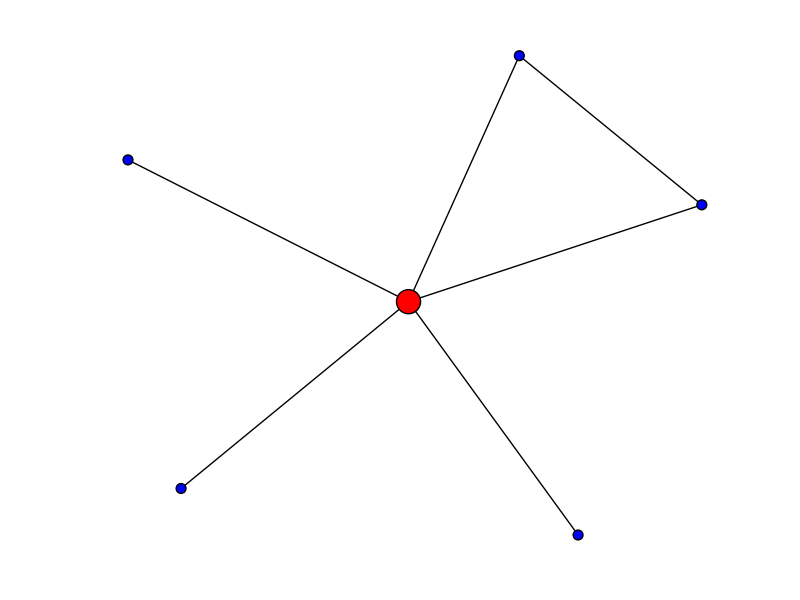}
   \label{fig:subfig6}
 }%
\end{subfigure}\hfill
 \begin{subfigure}[b][\small{Strongly linked (C7)}]{\includegraphics[trim = 0mm 0mm 0mm 10mm,scale=.30]{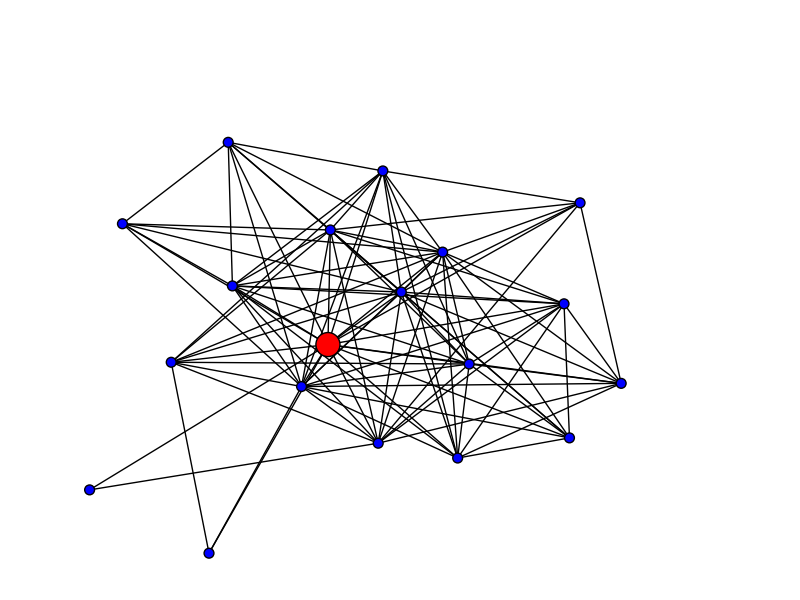}
   \label{fig:subfig7}
 }%
\end{subfigure}%
 \begin{subfigure}[b][\small{Complete (C8)}]{\includegraphics[trim = 50mm 0mm 0mm 10mm,scale=.30]{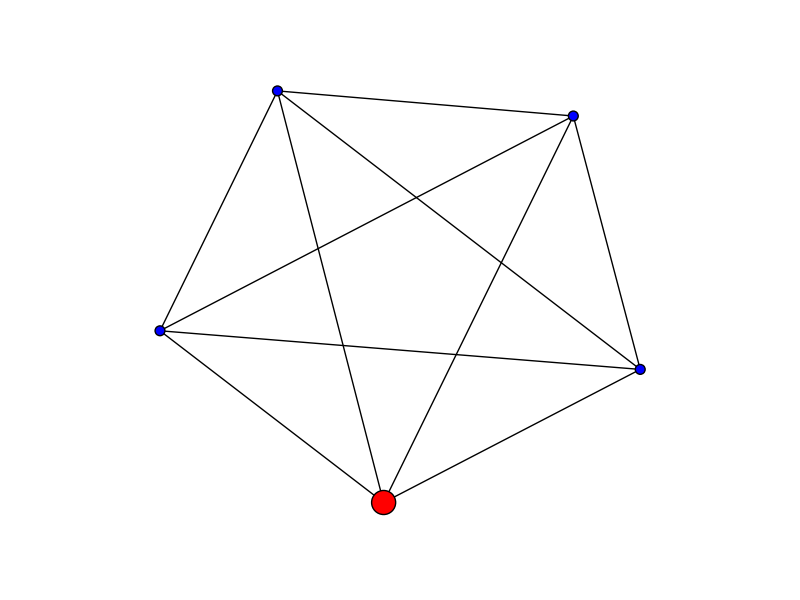}
   \label{fig:subfig8}
 }%
\end{subfigure}%
\label{fig78}
\caption{\small{Clustering results of ego graphs, depicting the ego node in the middle (red color), with connections to first- and second-degree neighbors. This study focuses on the automatic categorization of such ego graphs according to their graph structure. All in all, there exist eight distinct trends. We label each graph according to its characteristics.}}
\end{figure*}

{\bf Cluster(1) (Linked neighbors):} The ego node is the key player tied to active players, it is in a dense, active cluster at the center of events with many others. The structure has high closeness and low degree, betweenness centrality. The density of the ego graph is between 0.60 and 0.70. The ego node has many immediate neighbors. The neighbors of the ego are strongly connected to one another forming a strongly clustered network. The network has many complete structures and the second order neighborhoods are densely connected to each other.

{\bf Cluster(2) (Star):} Overall, the network has a sparse structure. The ego graph has a star structure and immediate neighbors of the ego are not connected. The number of neighbors varies depending upon the size of the network; the structure is small for smaller networks and large for bigger networks. The overall density of the ego graph is very low with no complete sub-graphs.  Normally, such clusters are identified by a high value of centrality measures. In such cases, we found that the strong second order neighborhood of the ego is more powerful and dense in terms of structures. 

{\bf Cluster(3) (Strong ego neighbors):} The ego node has few immediate neighbors and the network has some cohesive structures. The structure density is between 0.50 and 0.60. Some of the nodes in the network are highly populated and more powerful than the ego. The structure has higher closeness and degree, but low betweenness centrality.

{\bf Cluster(4) (Dense):} The ego node is an active player in the network and contains a fair amount of immediate neighbors. The neighbors of the ego node are well connected and have high density values between 0.70 and 0.80. Apart from very few neighbors (2 or 3), most of them form a strong cohesive network. The overall structure of the network is big with many complete sub-graphs. Even in case of removing the ego node from the network, it still contains many complete networks and the information can be easily transfered to other nodes. The ego's connections are highly redundant and most communication bypasses him. 

{\bf Cluster(5) (Powerful ego node):} The structure has high closeness and eigenvector centrality values and the ego has an average size neighborhood. The ego node is the most powerful player and removing it will paralyze the network. The structure has very few and small complete networks. Overall, the ego graph is not well connected with low graphs density. Normally, such nodes have few neighbors, but they act as boundary spanners. The ego node controls the communication between different parts of the network.

{\bf Cluster(6) (Less cohesive star):} Overall, the network contains small structures. The network has few nodes that are not well connected. It contains few undirected complete triads, and the remaining neighbors form a star-shape network. 

{\bf Cluster(7) (Strongly linked):} The structure is highly populated with many immediate neighbors and high internal density. The ego graph has many complete structures. The ego has a highly populated second order neighborhood. Many nodes within the graphs are populated. The sub-graphs have very high density in range of 0.80 and 0.90. There are multiple paths in the networks for the transfer of information. 

{\bf Cluster(8) (Complete):} The ego has few immediate neighbors, but they form a strong cluster. The structure is complete and the density of the network is 1.0. It shows that ego and its neighbors are actively in contact with each other.

\begin {table*}
\small
\begin{center}
\begin{tabular}{ | p{2cm} | p{4.75cm} | l | l | l | l | l | l | l|l|l|l|}
\hline
Datasets & Feature Spaces&  C1 & C2 & C3 &C4 &C5 &C6 &C7& C8&E&$H_R$ \\ \hline
\multirow{8}{20em}{Nokia GPS} & & &&&&&&&&& \\
& Centrality Measures &{+}{*}{!}{=}&  & &{+}{*}{!}  &&{+}{*}{!}{=}&{+}{*}{!} &{*}{!}{=}   &0.33&1.56 \\ \cline{2-12}
& Efficiency Measures & {+}{*}{!}&  & &{+}{*}{!} & &{+}{*}{!} && {+}{*} &0.47&1.29 \\ \cline{2-12}  
& Transitivity Measures &{+}{*}{!}& &  &{+}{*}{!} & &{+}{*}{!} && {+}{*}{!} &0.37& 1.38\\ \cline{2-12}
& Centrality and Efficiency & {+}{*}{!}& & & {+}{*}{!} &&&{+}{*}{!}&{+}{*}{!}  &0.51&0.96 \\ \cline{2-12}
& Centrality and Transitivity Measures &{+}{*}{!}&  & &{+}{*}!& &{+}{*}{!} &{+}{*}& {+}{*}!&0.29&1.90  \\ \cline{2-12}
& Efficiency and Transitivity Measures & {+}{*}{!}& & &{+}{*} &&{+}{*}{!} &{+}{*}{!} & {*}&0.69&1.15 \\\cline{2-12}
& Actor Based Measures & {+}{*}{!}&  &  &{*}{!}{+} & &&{+}{*}{!} &{+}{*}{!} &0.28&1.83  \\\cline{2-12}
& All Measures  & {+}{*}{!}& &  & & &&&&0.77&0.50 \\
\hline
\multirow{8}{20em}{Nokia Bluetooth} & & &&&&&&&&&  \\  
& Centrality Measures & {+}{*}{!}&  & &{+}{*}{!} & & &&&0.48&1.57  \\ \cline{2-12}
& Efficiency Measures & {+}{*}{!}& & & & &&&&0.63&0.87  \\ \cline{2-12}
& Transitivity Measures &{+}{*}{!}& & & & {+}{!}&{+}{*}{!} && &0.51&1.48 \\\cline{2-12}
& Centrality and Efficiency &{+}{*}{!}&  & & & &{*}{!} &&&0.47 &1.61 \\ \cline{2-12}
& Centrality and Transitivity Measures &{+}{*}{!}&  &&& &{+}{*}{!}&{+}{*}{!}& {+}{!}&0.43&1.67 \\ \cline{2-12}
& Efficiency and Transitivity Measures & {+}{*}{!}& & & & &{!} &+& {*}  &0.60&0.97  \\ \cline{2-12}
& Actor Based Measures &{+}{*}{!}&  &  & {+}{*}& &{+}{*}{!} &{+}{*}{!}& {+}{*}&0.39& 1.80 \\ \cline{2-12}
& All Measures  &{+}{*}{!}& & & & &{+}{!} &{+}{!}& {+} &0.67&0.91  \\
\hline
\multirow{8}{20em}{Social Evolution} & & &&&&&&&&& \\
& Centrality Measures & {+}{*}{!}& &&& & &{+}{*}{!}&& 0.23&0.48 \\  \cline{2-12}
& Efficiency Measures & {+}{*}{!}& & & & & &&&0.37&0.28   \\  \cline{2-12}
& Transitivity Measures &{+}{*}{!}&  &  &{+}{*}{!} && &{+}{*}{!}&*& 0.32&0.36 \\  \cline{2-12}
& Centrality and Efficiency & {+}{*}{!}& & & & & &&&0.29&0.39  \\  \cline{2-12}
& Centrality and Transitivity Measures &{*}&  & &{*}! & & &{+}{*}{!}&&0.34&0.31  \\ \cline{2-12}
& Efficiency and Transitivity Measures & {+}{*}{!}& & & & & &&&0.48&0.23 \\  \cline{2-12}
& Other Four Measures & {+}{*}{!} & & & & & &&&0.19&0.63 \\  \cline{2-12}
& All Measures  & {+}{*}{!} & & & & & &&&0.71&0.12\\
\hline

\multirow{10}{20em}{Friends and \\ Family} & & &&& &&&&&& \\
& Centrality Measures & {+}{*}{!}&  & &{+}{*}{!}& &{+}{*}{!} &&& 0.33&1.15 \\  \cline{2-12}
& Efficiency Measures &{+}{*}{!} & &  &{+}{*}{!} & &&&& 0.45&1.03  \\ \cline{2-12}
& Transitivity Measures &{+}{*}{!}&  & &{+}{*}{!}& & &&& 0.53&0.86 \\  \cline{2-12}
& Centrality and Efficiency & {+}{*}{!}&  & &{+}{*}{!} & & && &0.41&1.11 \\  \cline{2-12}
& Centrality and Transitivity Measures &{+}{*}{!}& & &{+}{*}{!}& &{+}{*}{!} && &0.41&1.11\\  \cline{2-12}
& Efficiency and Transitivity Measures & {+}{*}{!}& & & & &&&&0.47&0.98 \\ \cline{2-12}
& Other Four Measures & {*}&  & &{+}{*}{!}&{+}{*}{!}&{+}{*}{!} &&&0.63&0.74 \\  \cline{2-12}
& All Measures  & {+}{*}{!}& &  &{+}{*}{!}& &{+}{*}{!} &&&0.81&0.57   \\
\hline

\multirow{10}{20em}{Orange Dataset  } & & &&&&&&& && \\
& Centrality Measures & & {+}{*} &  & & &{+}{*}{!} &{*}& {+}{*}{!}&0.54&6.37 \\ \cline{2-12}
& Efficiency Measures & & {+}{*}{!} & & & &{+}{*} && {*}&0.83&3.11 \\ \cline{2-12}
&Transitivity Measures & & {+}{*}{!} &  & && &+{*}{!}& {+}{*}{!}&0.79&3.77 \\ \cline{2-12}
& Centrality and Efficiency & & {+}{*}{!} &  & &&{+}{*}{!} &{+}{*}{!}& {+}{*}{!}&0.64&4.98 \\ \cline{2-12}
& Centrality and Transitivity Measures &{+}{*}& {+}{*}{!} & {*} &{+}{*} & &{+}{*}{!} &{+}{*}{!}& {+}{*}{!}&0.71&4.23  \\ \cline{2-12}
& Efficiency and Transitivity Measures && {+}{*}{!} &  & & &{+}{*}{!} &{+}{*}{!}& {+}{!}&0.85&3.01 \\ \cline{2-12}
& Other Four Measures & & {+}{*}{!} & & & &{*}{!} &{+}{*}{!}&&0.49&6.78 \\ \cline{2-12}
& All Measures  & & {+}{*}{!} & & & &&&&0.89&2.77  \\
\hline

\end{tabular}
\caption {\small{Clustering results for four Datasets using four clustering algorithms. The first row represents the datasets. We use Bluetooth and GPS data from the Nokia mobile dataset. The second row represents the formed feature subspaces. Similarly, C1, C2,\ldots, C8 represent the short form of cluster 1, cluster 2,\ldots, cluster 8. The E and $H_R$ represent the entropy and representation entropy scores for the feature subsets. We use four signs to represent clustering algorithms. The +, * and !  signs represent the k-means, hierarchical clustering and affinity propagation (AP) respectively. There are some cells without entries representing that for some combinations that particular shape is not detected.} }
\label{table6}
\end{center}

\end{table*}

Table~\ref{table6} represents the clustering results for the four datasets. The rows and columns represent the datasets, formed feature spaces, and the clusters identified from the feature spaces. The +, * and ! signs represent the k-means, hierarchal clustering and affinity propagation respectively. We modeled Bluetooth and GPS modality from the Nokia dataset.

\begin{figure*}
\begin{subfigure}[b][\small{Dominant clusters for politically \\motivated participants}]{\includegraphics[trim = 20mm 0mm 0mm 0mm,scale=.20] {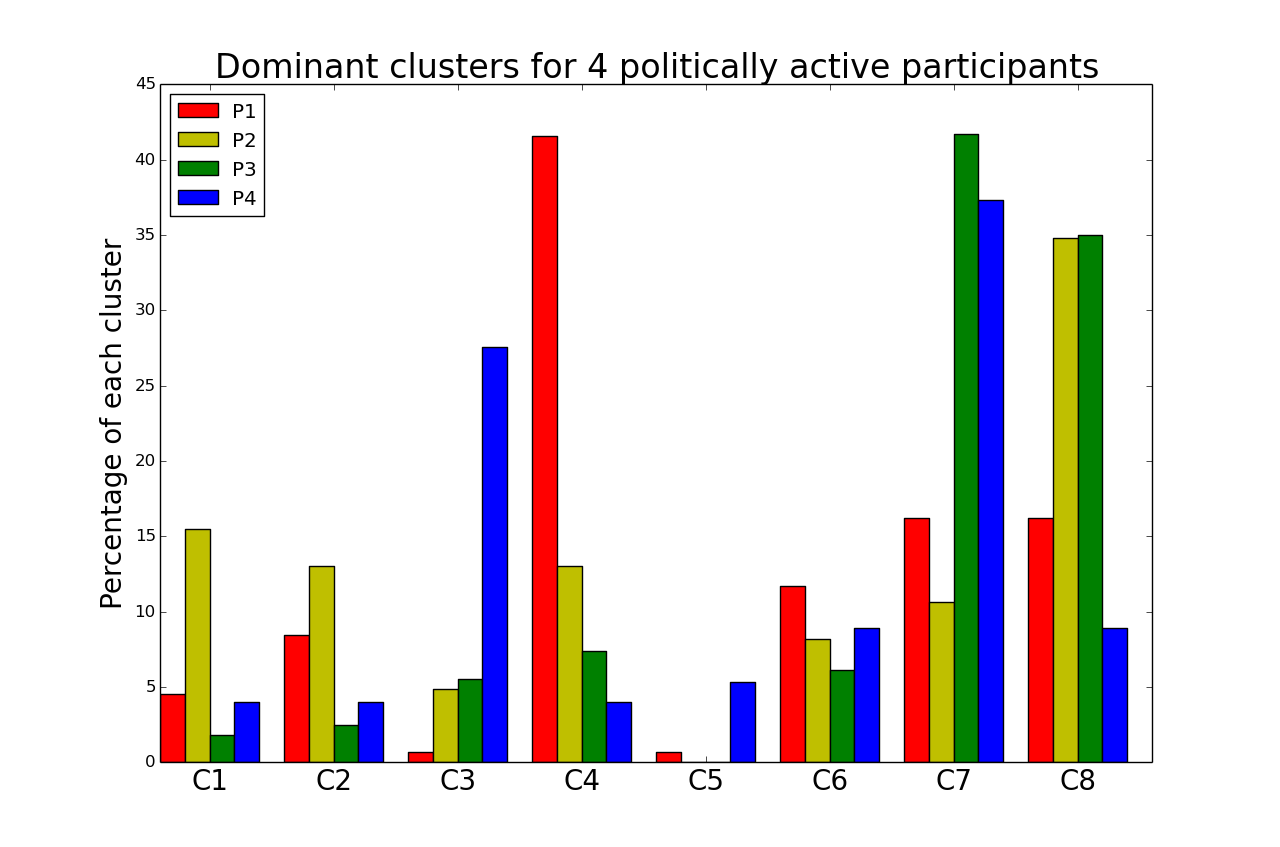}
   \label{polit}
 }%
\end{subfigure}
 \begin{subfigure}[b][\small{Dominant clusters for socially active\\ participants}]{\includegraphics[trim = 20mm 0mm 0mm 0mm,scale=.20]{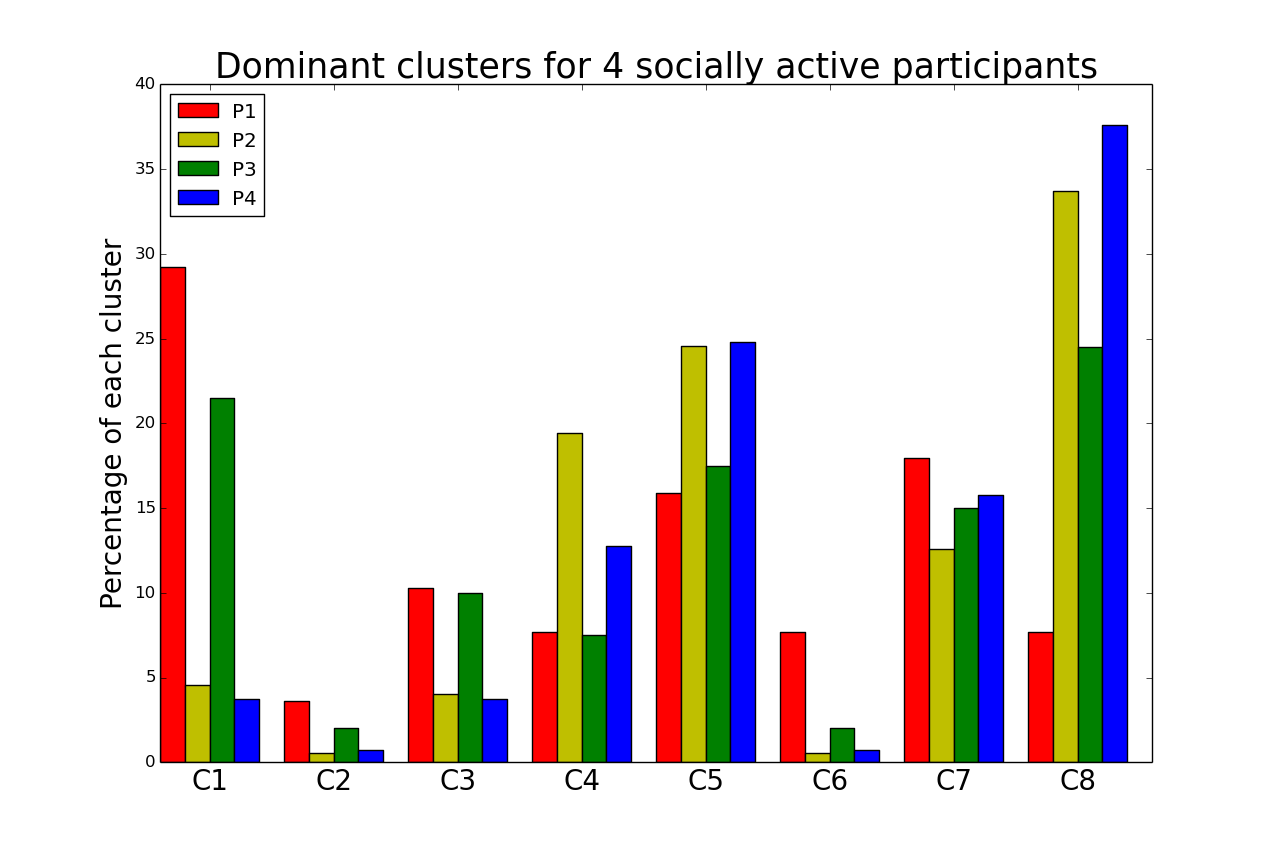}
   \label{social}
 }%
\end{subfigure}
 \begin{subfigure}[b][\small{Politically active participants}]{\includegraphics[trim = 25mm 20mm 0mm 0mm,scale=.35]{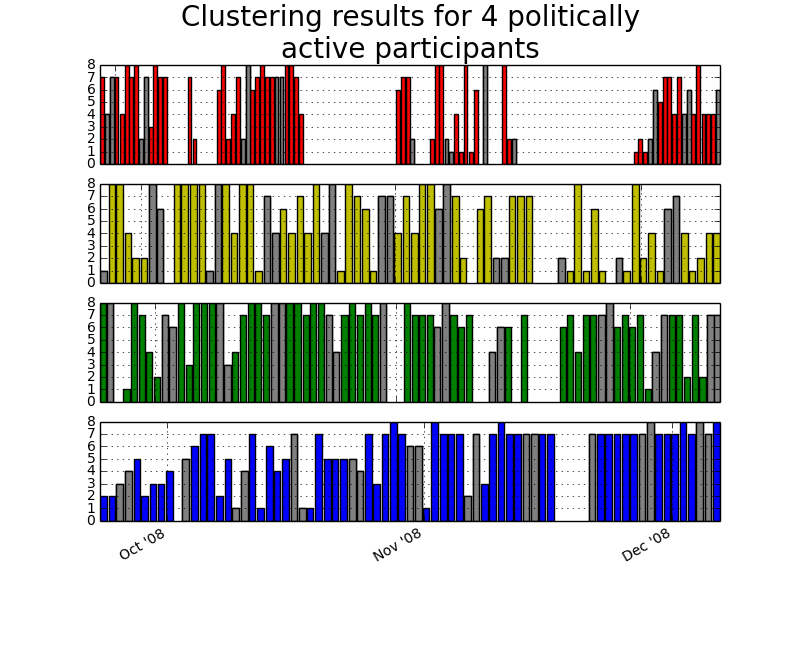}
   \label{ppp}
 }%
\end{subfigure}
\caption{\small{Clustering patterns for politically motivated participants and socially active participants over a nine months period. The different colors illustrate different participants. The plots in the first column illustrate the percentage of occupation of each cluster, that means how much time the participant spent in different clusters. In the second column, each bar represents one day of the month and shows which cluster the participant was in on that particular day. The gray bars represent the weekends.}}
\end{figure*}

The results show that extraction of possible clusters is largely dependent upon the size of sample and environment of data collection. The Social Evolution and the Friends \& Family datasets were collected in a certain environment (student dormitory and married graduate students living in a campus facility) with people well familiarized with each other. Their clustering results illustrate that only certain clustering patterns are prominent. For the Social Evolution dataset, people are mostly confined within C1, C4 and C7 that represent strongly clustered structures without any sparse structures. Similarly, the Friends \& Family dataset analysis shows that C1, C4 and C6 are prominent. The remaining patterns hardly exist in the data. The Orange dataset is gathered from a large sample living in a diverse environment who hardly know each other. Their results show a large diversity of all possible clustering patterns. Table~\ref{table6} shows that mostly the data is concentrated in C2, C6, C7 and C8, but feature space v) with a hierarchical clustering and k-means is able to detect other possible clustering patterns. 

Centrality and transitivity measures (feature set v) produced optimum distinct clustering patterns for the Nokia GPS and the Orange mobile datasets. For each clustering result, we visually inspect the clustering patterns. We analyzed the clustering patterns by again extracting its features. We found the best clustering results with a hierarchal clustering algorithms and k-means. However, the clustering results derived from k-means contained many misclassified clusters. We also detected misclassified clusters for hierarchical clustering, but they were very few. On the other side, we also found weak clustering results for some combinations. The combination of all features (feature space viii) produced poor results. It produced only one cluster for the Social Evolution and the Orange mobile datasets. Similarly, feature spaces ii), iv) and vi) also generated weaker results in many cases. Affinity propagation produced decent results on smaller datasets, but generated many outliers for bigger datasets. For feature space ii) in the Orange dataset, the number of clusters were equal to the size of sample. 
We found that feature spaces v) and vii) with a hierarchical clustering algorithm and occasionally k-means produced best clustering results. However, we need a better automated approach to select such features automatically. We select FSFS \cite{Mitra:2002:UFS:507475.507477} to automatically select feature subsets based on the characteristics of the data. For further evaluation, we select feature subsets from FSFS and then feed it as an input to hierarchical clustering algorithm. We study results using hierarchical clustering, because it produces best results for all four datasets. We studied two datasets over several months as a proof of concept for the analysis of such networks over time.  

\section{Case Studies}

In the previous section, we described the predictive power of different feature spaces and the clustering patterns detected by them. In this section, we analyze important clues from two datasets over an extended period of time. We discuss empirical results for ego networks based on ground-truth information. We focus our analyses on two aspects. First, we are interested to find the existence of any possible relationship between the clustering patterns and users' behavior. We merely rely on factual results without getting into any speculations. Second, we are interested to investigate the applicability of our designed tool over a long period. We study two clues from the Social Evolution dataset and two from the Friends and Family dataset. The ground-truth for the clues are provided with the datasets \cite{18,16}. It is important to note that our results are based on a relatively small sample size that might not be a representative of large real-world groups. However, they provide a starting point for the discussion on the study of distinctive ego neighborhood patterns. 

\subsection{Extraction of Clues from the Social Evolution Dataset}
We demonstrate the clustering patterns for the following two clues: i) politically motivated participants, ii) socially active participants.
\begin{figure*}
\begin{subfigure}[b][\small{Dominant clusters for two married couples}]{\includegraphics[trim = 20mm 0mm 0mm 20mm,scale=.32] {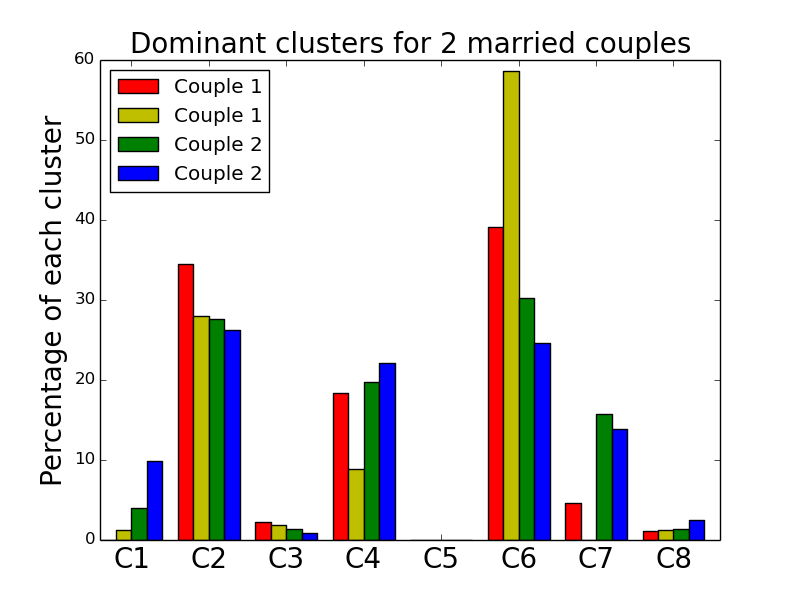}
   \label{marry}
 }%
\end{subfigure}
 \begin{subfigure}[b][\small{Dominant clusters for four friends}]{\includegraphics[trim = 20mm 0mm 20mm 20mm,scale=.32]{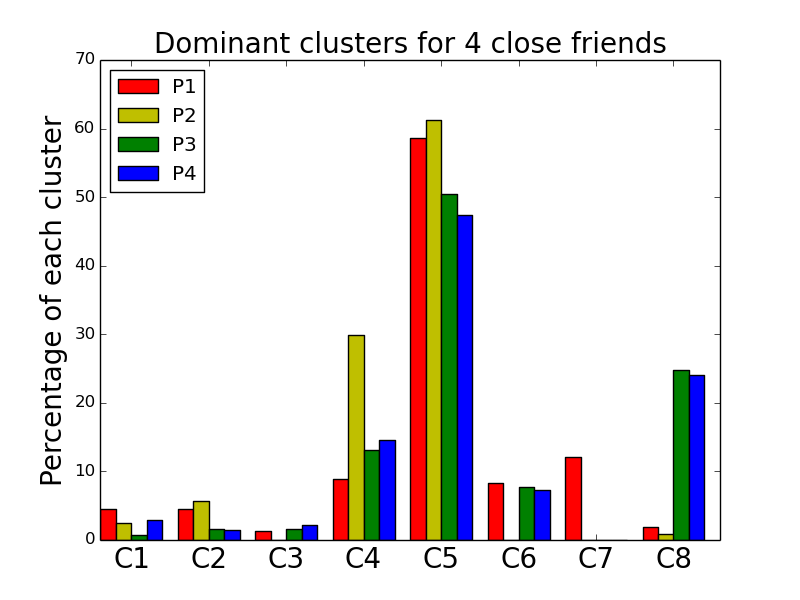}
   \label{closefriends_c1}
 }%
\end{subfigure}
 \begin{subfigure}[b][\small{Clustering rhythms for married couples}]{\includegraphics[trim = 0mm 20mm 0mm 0mm,scale=.37]{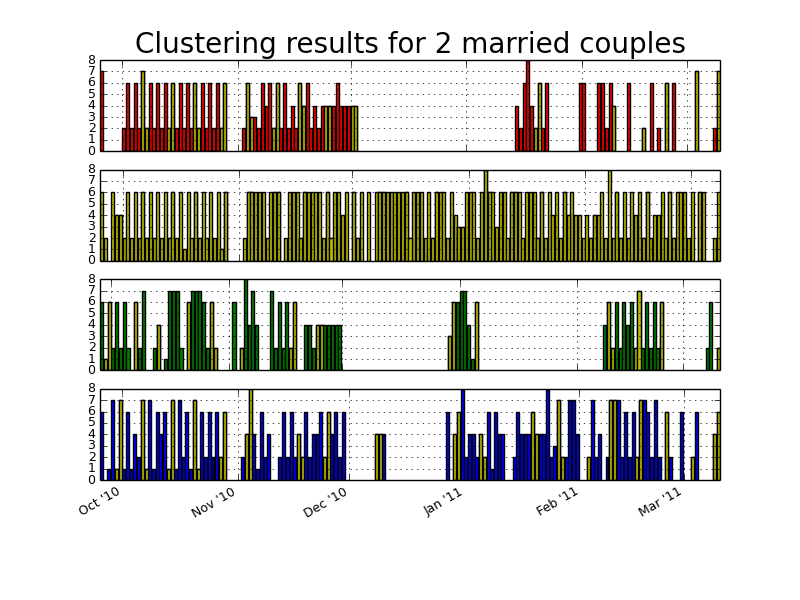}
   \label{marry_bar}
 }%
\end{subfigure}
\caption{\small{Illustrations for two married couples and four close friends. Figure~\ref{marry} shows C2, C4, C6 and C7 are the dominant clustering patterns for the married couples. In Figure~\ref{marry_bar} the red and yellow bars represent the first couple, and green and blue represent the second couple. Similarly, Figure~\ref{closefriends_c1} shows that C4 and C5 are the dominant clustering patterns for the close friends. }}
\end{figure*}
\subsubsection{Clustering Patterns for Politically Motivated Participants}
We examine the clustering patterns for the students during nine months of data collection. We found that self-reported political discussants have characteristic interaction patterns and this can be seen from their clustering patterns. Figure~\ref{polit} shows the clustering results for a sample of four participants that reported keen interest in politics. The red, yellow, green, and blue color represent four participants. It represents the percentage of time a user remained within a certain cluster. 
Figure~\ref{polit} shows the emerging clusters for participant 1 (C4, C7 and C8), participant 2 (C1, C4 and C8), participant 3 (C4, C7 and C8) and participant 4 (C3, C6 and C7). Mostly, the participants are in C4, C7 and C8 that are dense with strong neighborhood structures. Figure~\ref{ppp} shows the clustering patterns for a three months period; especially the focus is on the last 45 days of the 2008 US presidential election campaign. Participant 1, 2, and 3 reported consistent interest in politics; however, an interesting pattern can be observed for the fourth participant in Figure~\ref{ppp}. The clustering patterns vary for the first two halves of October, but then onwards show a more stable clustering pattern. In November, the fourth participant changes his political opinion from moderately interested to highly interested. For the remaining period, participant 4 was mostly in C3 and C7.
\subsubsection{Clustering Patterns for Socially Active Participants}
Figure~\ref{social} shows the clustering patterns for four socially active participants. The participants actively participated in campus activities and sports activities. The participants are mostly in C1, C3, C5 and C7. Figure~\ref{social} shows that participant 1 and 4 are around 25\% of the time in C5. In section V, we discussed that C5 acts as boundary spanner between different clusters, it acts as a bridge between different groups.  
\subsection{Extraction of Clues from the Friends and Family Dataset}
The dataset contains the information for 26 couples living in an undergraduate neighborhood. We demonstrate the clustering patterns for the following two clues: i) married couples, ii) close friends.
\subsubsection{Clustering Patterns for Married Couples}
We select two married couples to investigate their clustering patterns. Figure~\ref{marry}, and ~\ref{marry_bar} illustrate the clustering patterns for the married couples. The characteristic patterns for each couple have many similarities. For the first couple, C2, C4 and C6 are dominant, similarly for the second couple, C2, C4, C6 and C7 are dominant. We examine the data of all 26 couples to detect their clustering trends. The results suggest that the clustering patterns for married couples are mostly in C2 and C6. The overall structures were small and less cohesive. 
\subsubsection{Clustering Patterns for Close Friends}
We select four friends from the relationship survey that reported close acquaintance with each other. Figure~\ref{closefriends_c1} illustrates their clustering patterns. The sample does not contain any interaction data concerning people not participating in the data collection, the clustering patterns for four participants are very much alike. The detected structures are small, and mostly C4, C5 and C8 are detected.




\section{Conclusion} \label{conclusion}
In this report, we study so-called ego graphs extracted from four spatial-temporal datasets to characterize their neighborhood patterns. We consider two types of interactions as sensed by mobile phones, namely physical proximity (Bluetooth and GPS) and call-log data. The major contributions of the report are threefold:
\begin{itemize}
\item First, we examined in a systematic way a wide range of network features (in particular those addressing the properties of the ego networks) and unsupervised clustering algorithms to identify the prototypical ego network patterns. Our empirical results have shown that selecting the best combinations of feature subsets and clustering algorithms to determine the optimal number of neighborhood patterns is surprisingly intricate. In addition, in case of a bad separation between the clusters, clustering algorithms tend to produce outliers and redundant clusters that can be misleading. 
\item Second, our clustering analysis detected eight prototypical emerging clusters for ego networks, each of them characterized by particular characteristics. We assigned labels to these prototypical clusters based on their shapes and properties of the ego and its neighborhood.
\item Finally, we analyzed two spatial-temporal datasets over several months as a proof of concept. We explored different clues, such as the clustering patterns of unhealthy people and married couples, to study different characteristic patterns. Interestingly, our analysis showed some predominant clustering patterns for different clues. For instance, we detected isolated behaviors (C2 and C6) for people reporting to be depressed or stressed. Despite the small sample size of the study, we believe that this offers an illustration of how such data-driven tools can be used in behavioral sciences for graph interpretations.
\end{itemize}

The interested reader can find more information and thorough analysis in \cite{ess_2015}.

\bibliographystyle{plain}

\bibliography{sigproc}

\end{document}